\documentclass[aps,twocolumn,superscriptaddress,showpacs,preprintnumbers,amsmath,amssymb,floatfix]{revtex4-1}

\usepackage{graphicx,tabularx}
\usepackage{dcolumn}
\usepackage{bm}
\usepackage{nicefrac}

\usepackage{upgreek}
\renewcommand{\figurename}{Fig.}
\newcommand{\figurenames}{Figs.}
\newcommand{\eq}[1]{Eq.~(\ref{eq:#1})}
\newcommand{\eqs}[1]{Eqs.~(\ref{eq:#1})}

\begin{document}
\preprint{APS/123-QED}

\title{Functionalization of atomic force microscopy Akiyama tips for
  magnetic force microscopy measurements}

\author{Markus Stiller}
\email{markus@mstiller.org} 

\author{Jos{\'e}  Barzola-Quiquia} 
\author{Pablo D. Esquinazi} 
\affiliation{Division
  of Superconductivity and Magnetism, Felix-Bloch Institute for
  Solid-state Physics, University of Leipzig, 04103 Leipzig, Germany}

\author{Soraya Sangiao}
 \affiliation{Laboratorio de Microscop{\'i}as
  Avanzadas (LMA), Instituto de Nanociencia de Arag{\'o}n (INA),
  Universidad de Zaragoza, E-50018 Zaragoza, Spain}
\affiliation{Departamento de F{\'i}sica de la Materia Condensada,
  Universidad de Zaragoza, E-50009 Zaragoza, Spain} 

\author{Jos{\'e}  M. De Teresa} 
\affiliation{3' Instituto de Ciencia de Materiales de
  Arag{\'o}n, CSIC-Universidad de Zaragoza, 50009 Zaragoza, Spain}
\affiliation{Laboratorio de Microscop{\'i}as Avanzadas (LMA),
  Instituto de Nanociencia de Arag{\'o}n (INA), Universidad de
  Zaragoza, E-50018 Zaragoza, Spain} \affiliation{Departamento de
  F{\'i}sica de la Materia Condensada, Universidad de Zaragoza,
  E-50009 Zaragoza, Spain} 

\author{Jan Meijer} \affiliation{Division
  of Nuclear Solid State Physics, Felix-Bloch Institute for
  Solid-state Physics, University of Leipzig, 04103 Leipzig, Germany}

\author{Bernd Abel} \affiliation{Leibniz Institute of Surface
  Modification, 04318 Leipzig, Germany}
		
\date{\today}

\begin{abstract}
  In this work we have used focused electron beam induced deposition
  of cobalt to functionalize atomic force microscopy Akiyama tips for
  application in magnetic force microscopy.  The grown tips have a
  content of $\approx 90~\%$ Co after exposure to ambient air.  The
  magnetic tips were characterized using energy dispersive X-ray
  spectroscopy and scanning electron microscopy. In order to
  investigate the magnetic properties, current loops were prepared by
  electron beam lithography. Measurements at room temperature as well
  as $4.2~\mathrm{K}$ were carried out and the coercive field of
  $\approx68$~Oe of the Co tip was estimated by applying several
  external fields in the opposite direction of the tip magnetization.
  Magnetic Akiyama tips open new possibilities for wide-range
  temperature magnetic force microscopy measurements.
\end{abstract}

\maketitle


\section{Introduction}
\label{introduction}

Magnetic force microscopy (MFM) is an essential tool to probe the
local stray fields of materials and can access information from
nanometer to micrometer length scales. This is important for
experimental work in the area of nanotechnology, where MFM offers
superior spatial resolution compared to superconducting quantum
interference measurements~\cite{KKSSGBW95}. Thus, MFM has been used to
investigate and to manipulate vortices in
superconductors~\cite{MHSG98,VTBHMSDHG02,ALSEKZBLHM08} or
ferromagnetic domain walls in thin films~\cite{HPOB96,SRBC95,GTM01}
and micro/nano-structures~\cite{CBKSHJPLS10}. However, most MFM
systems are restricted to room-temperature measurements, lacking the
possibility to explore the temperature dependence of magnetic and/or
superconducting properties in such samples.  Another crucial factor
is, that conventionally AFM devices are operated using a laser to
detect the cantilever oscillation, which implies a working place of
the order of tens of centimeters, while for AFM systems implemented
with Akiyama tips, the working place is reduced to a few centimeters,
making it possible to place them inside a cryostat or employ
multi-probe scans. Furthermore, these tips can be used in darkness and
thus allow for measurements inside tubes or a combination of AFM$/$MFM
with scanning electron, optical or confocal microscopy.  An example of
Akiyama tips used for magnetic measurements is the combination with
diamonds containing NV-centers which are glued at the end of the
tip.~\cite{RTSSKD12} In addition, low-temperature measurements are
useful to rule out topographic or electrostatic
contributions~\cite{SMPFCSRCS17}. Often, samples undergo a variety of
phase transitions upon cooling, such as magnetic quantum tunneling or
magnetic anisotropies of small particles that exhibit
superparamagnetism at room temperature. The hardware for
self-oscillation and self-sensing tips requires less space and can
be placed easily in a cryostat for temperature dependent
measurements. Low-temperature MFM has been used to study micrometric
aggregates of a paramagnetic gadolinium acetate
complex~\cite{LJMASMRE13} or magnetic switching of ferromagnetic
Prussian blue analogue nanoparticles~\cite{CVAC16} or single molecule
magnets~\cite{SMPFCSRCS17}.

Moreover, the MFM device itself profits from low temperature, since
thermally induced drifts and piezo creeps are smaller, which is very
useful for devices where closed-loop scanning is not possible. The
cryo-cooled AFM device used for this work, was operated using Akiyama
tips, for which no magnetic tips are commercially available. In order
to produce magnetic tips, focused electron beam induced deposition
(FEBID) of Co was used. Current loops were prepared to characterize
the tips and also for comparison using standard MFM tips with a
standard AFM/MFM device. Using detailed theory and computing the
equations with standard programs, simulations were done to compare
with the results.

\section{Experimental Procedure}
\begin{figure*}
\includegraphics[width=\textwidth]{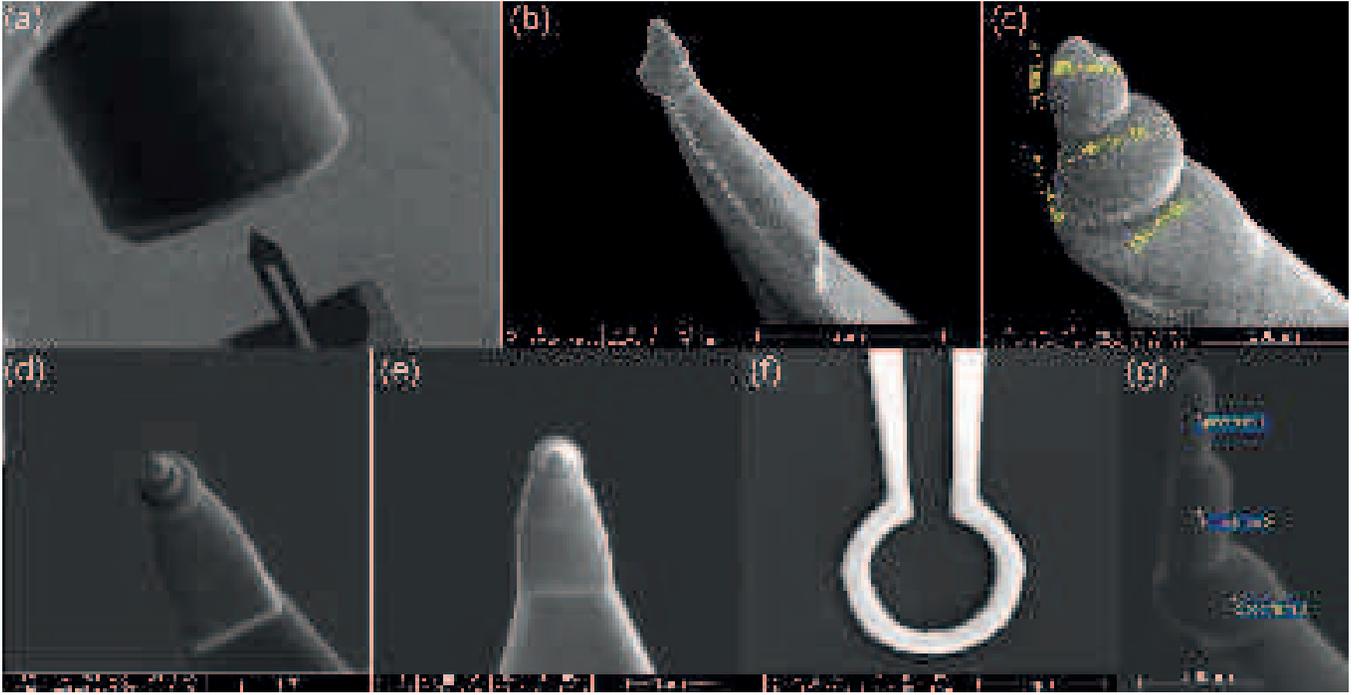}
\caption{\label{fig:fig1} The Akiyama cantilever and gas injection
  system is shown in (a). A deposited magnetic tip can be seen in (b)
  and (c). Pictures (d) and (e) show the magnetic tip before (d) and
  after measurements (e). The current loop to investigate the response
  on a magnetic field can be seen in (f), and the crosses in (g)
  indicate where the EDX spectra were recorded.}
\end{figure*}

For the experiments commercial Akiyama tips (Nanosensors) designed for
AFM were used, with a spring constant of $k=5$~N$/$m.  For comparison,
a conventional MFM device (Veeco) with standard MFM tips (Bruker,
$k=3$~N$/$m, Q=220) was used. The measurements with Akiyama tips were
done using a AFM/CFM device (Attocube) placed in a He-cooled cryostat
(Oxford), low-temperature measurements had to be performed using a
phase-locked loop due to the large qualify factor $Q=2757$ at
$T=4.2$~K. Magnetic Akiyama tips have been prepared by FEBID using a
dual-beam Helios 600 system (FEI) that integrates an electron beam and
an ion column, see \figurenames~\ref{fig:fig1}((a)--(e)). FEBID is an
\textit{in-situ} deposition method, allowing the production of
nano-to-micrometer-sized structures with desired thicknesses and
shape.

During the deposition, the focused electron beam current was set to
0.17~nA at 5~kV acceleration voltage and a $\rm Co_2 (CO)_8$ gas
precursor was injected. Precursor molecules, delivered onto the
substrate surface by means of a nearby gas-injection system, are
dissociated by the electron beam irradiation, giving rise to a deposit
with the same shape of the beam scanning~\cite{TPCRSI16}. FEBID allows
the precise growth of a magnetic tip at the apex of a cantilever with
controlled tip dimensions~\cite{UHBS02,LNKLCT14}. In the present case,
a continuous electron irradiation on a single spot has been used for
the growth of the magnetic tip at the apex of the cantilever, see
\figurenames~\ref{fig:fig1}((b)--(e)).  Energy-dispersive X-ray (EDX)
spectra were acquired after exposing the cantilevers to ambient
conditions for several minutes, which gives rise to a saturated
surface oxidation layer of $\approx5$~nm~\cite{NMT16}, see
\figurename~\ref{fig:fig1}(g).

The tips were grown as follows: first a square-like structure was
deposited as basis for the tip, followed by two circles, with radii
300~nm and 100~nm. The resulting tip size was $\approx250$~nm, see
\figurenames~\ref{fig:fig1}((a)--(d)), due to the broadening effect
characteristic of the growth by FEBID~\cite{ATFOKRP14}. In
\figurename~\ref{fig:fig1}(e) the tip after measurement can be inspected,
and compared with \figurename~\ref{fig:fig1}(d), it is obvious that
deposited material is robust and suitable for AFM/MFM measurements.

The current loops were patterned with electron beam lithography, and
Cr$/$Au with a thickness of
\text{$\approx7$~nm$/$$\approx200$~nm,} was sputtered, see
\figurename~\ref{fig:fig1}(f). The loops used at room temperature have
a radius of $4~\upmu$m and $\approx
150$~nm width, whereas the current loops used at low temperatures have
a radius of $2.4~\upmu$m and a width of $\approx1~\upmu$~m.

\section{Results and Discussion}

Using EDX, the Co purity was determined to be between 85
and 92~at~\%., see Table~\ref{tab:t1}.
\begin{table}
\begin{tabularx}{\columnwidth}{XXXX}
  \hline\hline
  Spectrum &  C (At\%)& O (At\%)& Co (At\%)\\
  \hline
  1 & 6.88 & 1.31 & 91.81\\
  2 & 7.42 & 0.83 & 91.75\\
  3 & 12.23& 1.54 & 86.23\\
mean& 8.84 & 0.83 & 86.23\\
  \hline\hline
\end{tabularx}
\caption{EDX results of a magnetic Akiyama tip at positions shown in \figurename~\ref{fig:fig1}(g). The spectra were recorded after exposure to ambient air.}
\label{tab:t1}
\end{table}
For such Co content, electron holography measurements in Co nanowires
indicate that the magnetization along the long wire axis is around
$1$~T~\cite{NMT16}.
\begin{figure}
\includegraphics[width=\columnwidth]{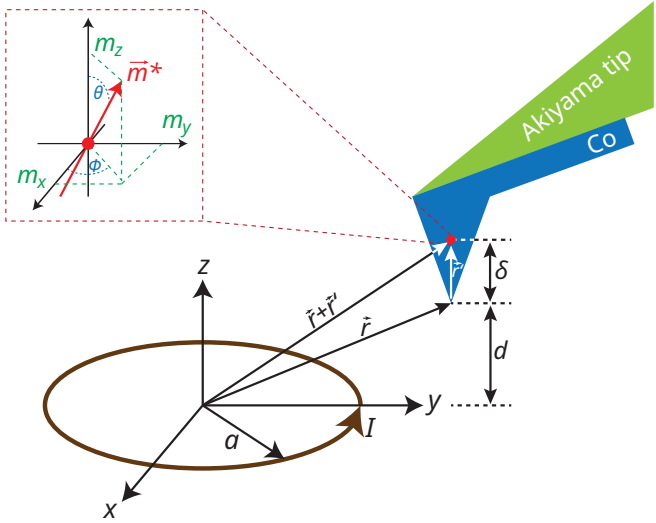}
\caption{\label{fig:fig2} Geometry of current loop and Akiyama tip
  with deposited Co. The effective magnetic moment $m^*$ of the tip is
  treated as dipole with arbitrary angle.}
\end{figure}

In order to quantify the results, current loops were used to generate magnetic
fields. Considering a current loop of radius $a$, see
\figurenames~\ref{fig:fig1}(f) and ~\ref{fig:fig2}, located in the
$x-y$-plane, centered at origin and carrying a current $I$, the generated
magnetic field in Cartesian coordinates is~\cite{SLIY01}:
\begin{equation}
  \begin{aligned}
    \label{eq:Bfield}
    B_x&=\frac{C\cdot x\cdot z}{2\alpha^2\beta\rho^2}\left[\left(a^2+r^2\right)E(k^2)-\alpha^2K(k^2)\right],\\
    B_y&=\frac{C\cdot y\cdot z}{2\alpha^2\beta\rho^2}\left[\left(a^2+r^2\right)E(k^2)-\alpha^2K(k^2)\right] \\ &= \frac{y}{x}B_x,\\
    B_z&=\frac{C}{2\alpha^2\beta}\left[\left(a^2-r^2\right)E(k^2)+\alpha^2K(k^2)\right],\\
  \end{aligned}
\end{equation}
where $\rho^2=x^2+y^2$, $r^2=x^2+y^2+z^2$, $\alpha^2=a^2+r^2-2a\rho$,
$\beta^2=a^2+r^2+2a\rho$, $k^2=1-\alpha^2/\beta^2$, $C=\upmu_0I/\pi$,
$K$ and $E$ are the elliptic integrals of first and second kind,
respectively. The magnetic force derivative for an arbitrary
cantilever orientation is given by~\cite{WG92}
\begin{equation}
\begin{aligned}
  \label{eq:Fder}
  F'(\vec r)&=\int_{\rm tip}\sum_{i=x,y,z}\sum_{j=x,y,z}\sum_{k=x,y,z}n_jn_kM^T_i(\vec r')
  \\ &\times\frac{\partial^2B_i(\vec r + \vec r')}{\partial r_j\partial r_k}dV',
\end{aligned}
\end{equation}
where $\vec M^T$ denotes the tip magnetization. Assuming that the
sample magnetization and $\vec M^T$ are independent of each other,
that the cantilever is parallel to the sample surface,
i.e.~$\hat n=\hat z$, and that tip is a point dipole, one finds that
\begin{equation}
  \begin{aligned}
  \label{eq:Fder2}
   F'(\vec r)&=m_x\frac{\partial^2B_x}{\partial^2_z}+m_y\frac{\partial^2B_y}{\partial^2_z}+m_z\frac{\partial^2B_z}{\partial^2_z},\quad \mathrm{with}\\
\vec m^*&=\begin{pmatrix}m_x\\m_y\\m_z\end{pmatrix},
  \end{aligned}
\end{equation}
where $m^*$ is the effective magnetic moment of the deposited tip.
This provides the possibility to include an arbitrary angle of the tip
point dipole into the simulations, see \figurename~\ref{fig:fig2};
\textit{Matlab}\textsuperscript{TM} was used to calculate the derivatives of
\eqs{Bfield}. The phase shift of the cantilever can then be
simulated with
\begin{equation}
  \label{eq:dphi1}
  \Updelta\varphi=-\frac{Q}{k}\left(\frac{\partial F}{\partial z}\right)=\frac{2Q}{\omega_0}\Delta\omega,
\end{equation}
where $\omega_0$ and $\Delta\omega$ are the resonance frequency and
the frequency shift, respectively.  Assuming that $\vec M^T$ is
perfectly aligned parallel to $\hat z$, i.e.~$m_y=m_x=0$, then the
phase shift along the $z$-axis of the loop is given as:
\begin{equation}
  \label{eq:dphi2}
  \Updelta\varphi_z=\frac{3\upmu_0a^2Im_zQ}{2k}\left[\frac{a^2-4(d+\delta)^2}{(a^2+(d+\delta)^2)^{7/2}}\right],
\end{equation}
with $z=d+\delta$, where $d$ is the scan height and $\delta$ is the
distance between the Co tip and the center of the magnetic dipole. This
phase shift (\eq{dphi2}) is the difference of the phase between the
center of the ring and the phase signal at a large enough distance
away from the current ring with no stray fields present (as indicated
in \figurename~\ref{fig:fig4}(e)). This method has the advantage that
the magnetization will not be changed when scanning with the magnetic
tip or when applying an external magnetic field, which could be the
case when using a ferromagnetic sample. Also a linear current wire
could be used. However, this is not accurate because the MFM image
would have to be analyzed above an edge of the wire, in order to get
the $z$-component of the tip stray fields~\cite{CLKW01}. This not only
implies possible topographic influence but also electrostatic
distortion might occur due to contact potential between tip and the
current carrying wire~\cite{CLKW01}, which was seen clearly in the
measurements using the Akiyama tip. This could be avoided when
covering the wire with an insulating film and a gold film on top,
which then can be connected to the tip in order to electrostatically
shield the tip~\cite{BESAD96}.

\begin{figure}
\includegraphics[width=\columnwidth]{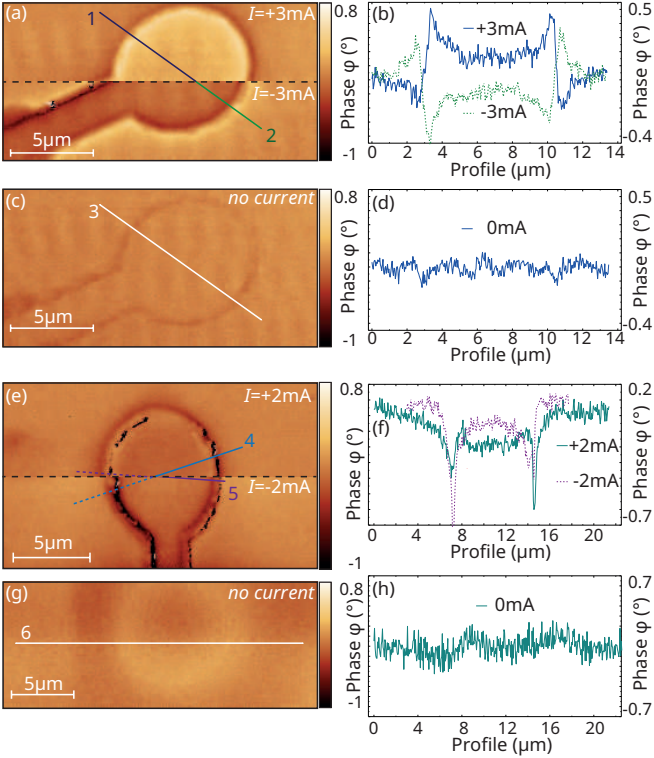}
\caption{\label{fig:fig3} Magnetic force microscopy images at room
  temperature of a current loop using a standard MFM tip and standard
  device for applied currents of $I=\pm3$~mA(a) and no applied
  current(c). Line scans as indicated can be seen in ((b)\&(d)).  The
  results using an Akiyama tip with applied currents of $I=\pm2$~mA
  are shown in (e), with no current applied in (g). The corresponding
  line scans can be found in ((f)\&(h)).}
\end{figure}

\begin{figure}
\includegraphics[width=\columnwidth]{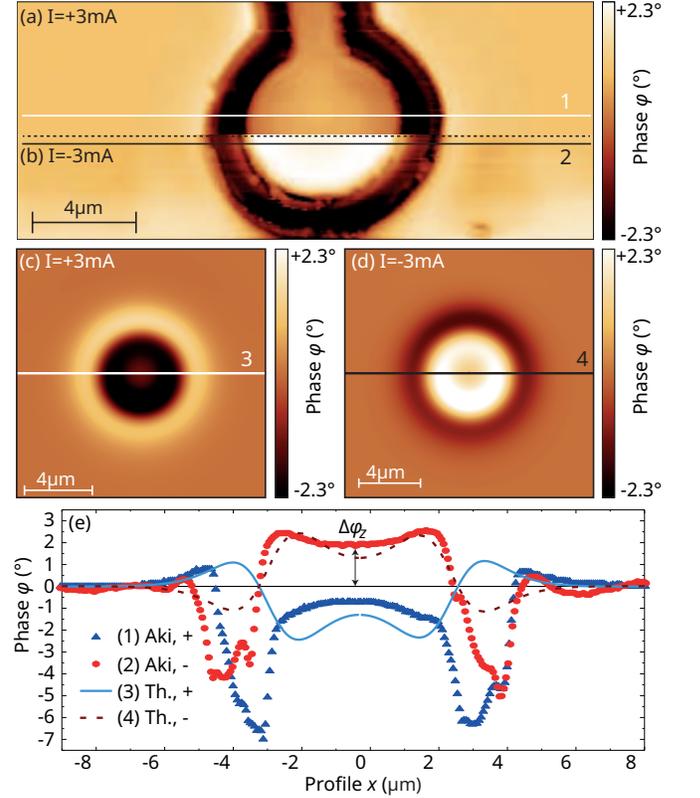}
\caption{\label{fig:fig4} MFM images of a current loop using an
  Akiyama tip at $T=4.2$~K, with applied currents of $I=+3$~mA (a) and
  $I=-3$~mA (b) and a scan height of $d=100$~nm. ((c) and (d)) the
  corresponding simulations, and the four spectra in (e), as
  indicated in the images.}
\end{figure}

In \figurename~\ref{fig:fig3} the results of MFM measurements at room
temperature are shown. The commercial tip together with the standard
AFM/MFM device yield expected results, see
\figurename~\ref{fig:fig3}((a)--(d)).
In \figurename~\ref{fig:fig3}(a) the top MFM image shows the results
when a current of $I=+3$~mA is applied, the bottom image was measured
with a current of $I=-3$~mA. The magnetic field at the center of the
loop was $B_c\approx\pm471~\upmu$T. The scan height of 90~nm has only
a low influence on the field strength, as the $1/e$ magnetic decay
length for a current loop is roughly equal to the
radius~\cite{LKCDW99}.  The two lines indicate the spectra shown in
\figurename~\ref{fig:fig3}(b), i.e.~spectrum 1 for positive and
spectrum 2 for a negative applied current, respectively. The phase
shift $\Updelta\varphi_z$ was found to be $\pm0.15^\circ$. In
\figurename~\ref{fig:fig3}(c) and (d) the results for zero current
are shown. Besides small topographic effects when scanning across the
wire, no signal is found at the center of the loop.

The results for Akiyama tip at room temperature can be seen in
\figurename~\ref{fig:fig3}((e)-(h)), with applied currents of
\text{$I=\pm2$~mA} resulting in magnetic fields of
\text{$B_c\approx\pm314~\upmu$T}. The phase shift for positive applied
current is \text{$\Updelta\varphi_{z,+}=-0.22^\circ$,} and for a
negative current, \text{$\Updelta\varphi_{z,-}=-0.14^\circ$},
respectively. This shows that the electrostatic interaction play an
important role when using Akiyama tips for MFM, when the tip and the
cantilever are electrically connected to the tuning fork. This becomes
even more evident when taking the MFM images into account, where large
parts of the loop show black dots, where the tip has struck the gold
surface. The phase at the center is also influenced by this effect,
because the Akiyama tip is tilted, as can be seen in
\figurename~\ref{fig:fig1}(b), and thus, even when the magnetic tip is
in the center of the loop, part of the tip is above the current
carrying Au wire. Topographic effects can be ruled out as source of
the negative phase shift, since there is no significant signal when
measuring with no applied current, see \figurenames~\ref{fig:fig3}(g)
and (h).

A more detailed characterization was possible at \text{$T=4.2$~K.} The
MFM images of a current loop can be seen in
\figurename~\ref{fig:fig4}(a) and (b),
for applied currents of \text{$I=\pm3$~mA}, the field at the center of
the loop is \text{$B_c\approx783~\upmu$T}, at a scan height of
$d=100$~nm. \figurenames~\ref{fig:fig4}(c) and (d) show the
corresponding simulations using \eq{dphi1}, the parameters $m_z$ and
$\delta$ are obtained through fit of $\Updelta\varphi_z(d)$ and
$\Updelta\varphi_z(I)$ to \eq{dphi2}, as is explained below. The four
spectra, as indicated in \figurenames~\ref{fig:fig4}((a)--(d)), are
shown in \figurename~\ref{fig:fig4}(e).  The phase shift for positive
applied current is \text{$\Updelta\varphi_{z,+}=-0.7^\circ$,} and for
a negative current, \text{$\Updelta\varphi_{z,-}=1.9^\circ$.} The
simulations are shown as (dotted) lines. For both applied currents,
experiment and theory agree well. Deviations are obvious when
measuring across the Au wire. This is due to the width of the wire of
$\approx1~\upmu$m, which is not taken into account in the simulations
(a perfect current loop with zero width). Furthermore, the tip has a
finite size and is not a perfect magnetic dipole. Both cause a
broadening of the signal compared to theory. Also, a large
negative contribution to the phase is evident when scanning above the
gold. This is due to the electrostatic interaction between the current
carrying loop and the Akiyama tip, when the tip hits the surface of
the gold wire.  The measurement of zero applied current did
not show any significant signal (not shown here), therefore
topographic effects can be ruled out as source of the negative phase.
The phase shifts at the center of the loop show a larger value for
both currents, when compared to the predicted one, see
\figurename~\ref{fig:fig4}(e). This indicates an additional repulsive
force ($\Updelta\varphi>0$ and $F'<0$). The sum
\text{$\Updelta\varphi_{z,\pm}=|\Updelta\varphi_{z,+}|+|\Updelta\varphi_{z,+}|=2.6^\circ$}. This
agrees with the simulations, which predicts
$\Updelta\varphi_{z,\pm}^{th}=2.6^\circ$.  Further, based on
\figurenames~\ref{fig:fig4}(a) and (b), an angle of the tip
magnetization was taken into account, i.e.~$\theta\approx17^\circ$ and
$\phi\approx81^\circ$, as defined in \figurename~\ref{fig:fig2}. This
yields a total tip magnetic moment of
$|m^*|=2.42\times10^{-11}$~Am$^2$.

\begin{figure}
\includegraphics[width=\columnwidth]{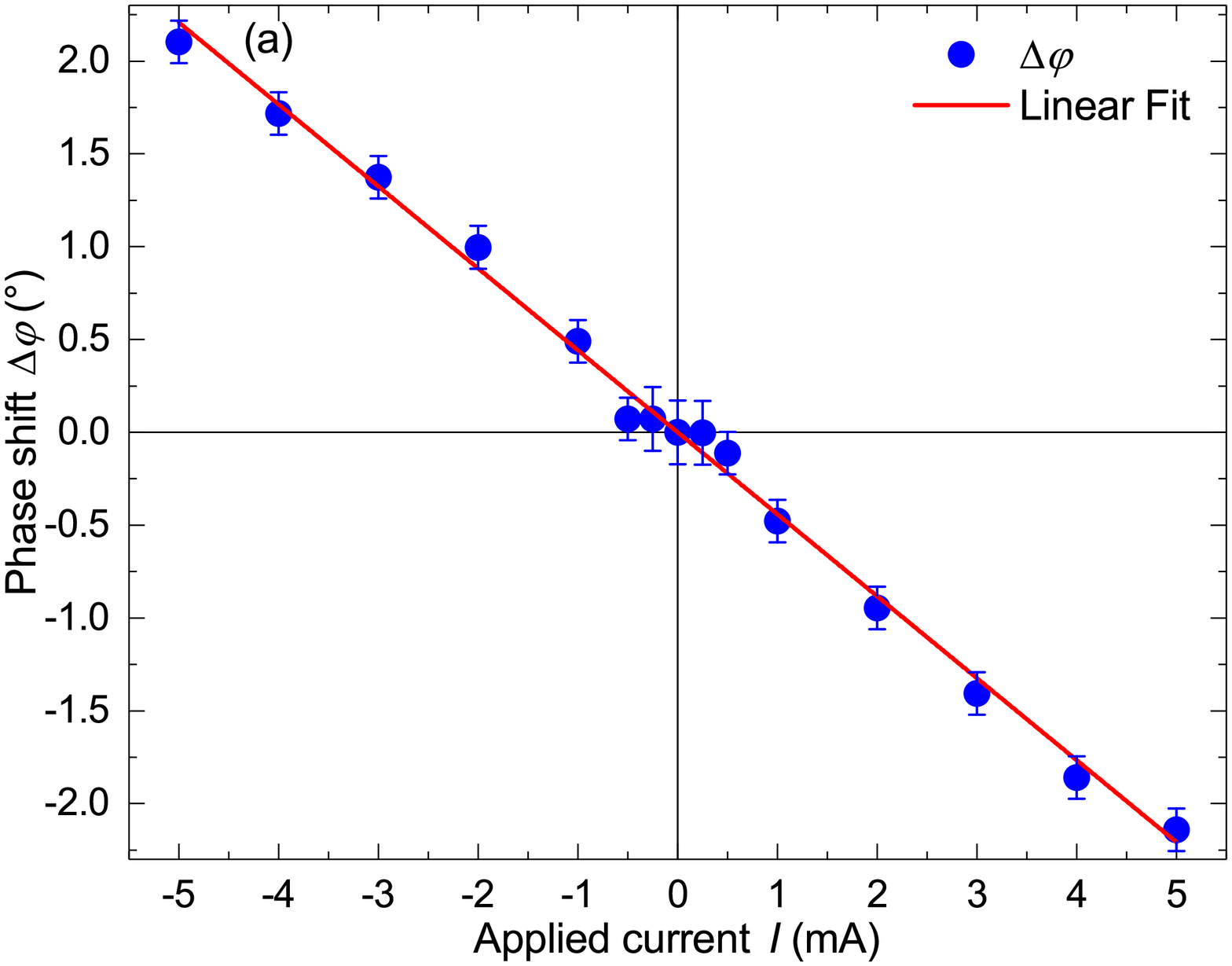}
\includegraphics[width=\columnwidth]{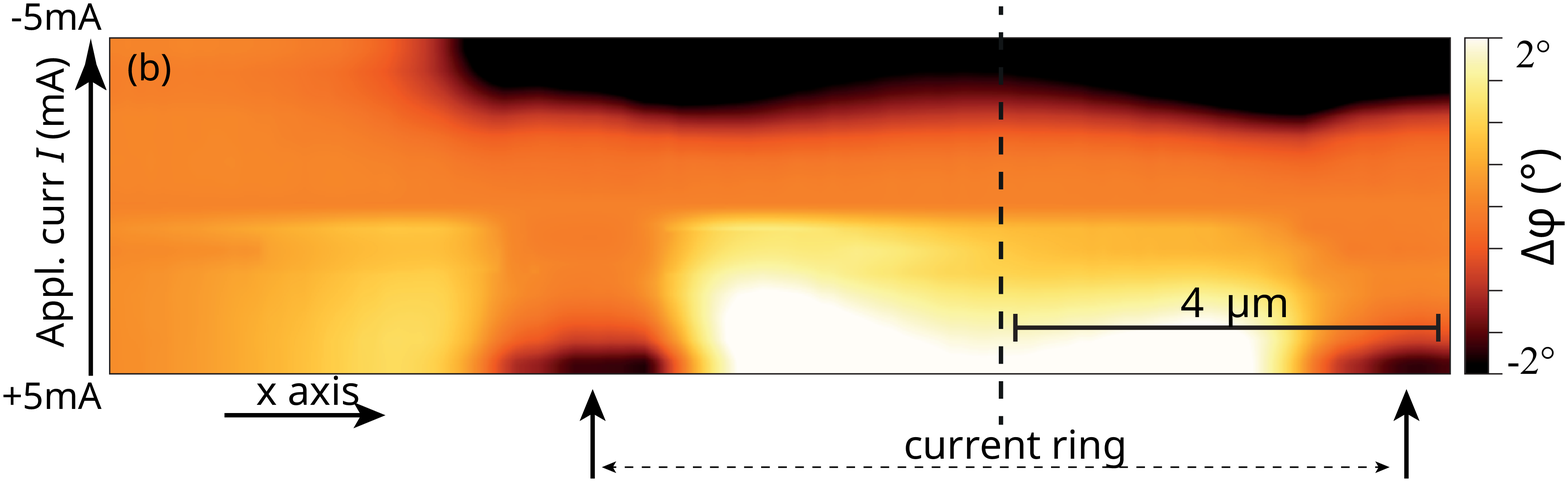}
\caption{\label{fig:fig5} (a) Phase shift $\Updelta\varphi$ as function of
  applied current at $T=4.2$~K. The red line is the fit to \eq{dphi2}. In (b) $\Updelta\varphi$ vs.~$I$ can be seen for a fixed $y$-scan axis.}
\end{figure}

\begin{figure}
\includegraphics[width=\columnwidth]{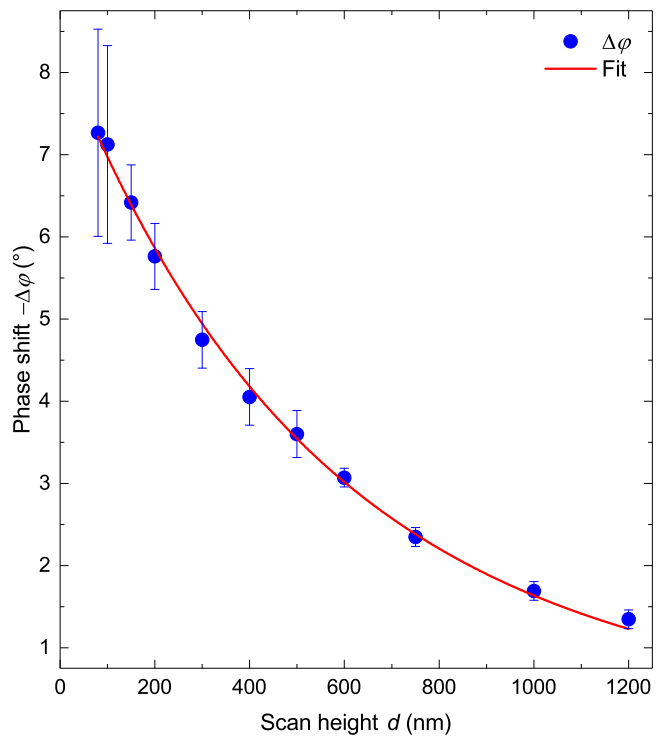}
\caption{\label{fig:fig6} Phase shift $\Updelta\varphi$ as function of
  scan height $d$ at \text{$T=4.2$~K}. The red line is the fit to
  \eq{dphi2}.}
\end{figure}

Using \eq{dphi2}, we can get information about the effective magnetic
moment of the tip in $z$-direction, $m_z$, and the dipole distance
from the tip peak, $\delta$. For this purpose, the $y$-scan axis was
fixed at zero and the center line was scanned several times. The
resulting spectra were averaged, in order to minimize the
noise. First, the phase shift at the center of the loop was measured
as function of applied current, see \figurename~\ref{fig:fig5}(a). The
positive offset was corrected.
As is expected from \eq{dphi2}, $\Updelta\varphi(I)$ is
linear, the straight red line is the fit. The current dependence can
also be seen in the MFM image in \figurename~\ref{fig:fig5}(b). There,
the $y$-scan axis was fixed at $y=0$ and the current was sweeped from
$I=+5$~mA to $I=-5$~mA. Similarly, the phase shift can be measured as
a function of scan height as is shown in \figurename~\ref{fig:fig6}.
The red line is the fit to \eq{dphi2}. From the fits of the current
and height dependence, an effective magnetic moment for the tip in
$z$-direction $m_z=2.39\times10^{-11}$~Am$^2$ and dipole distance
$\delta=1.85~\upmu$m were obtained for the current loop with radius
$a=2.4~\upmu$m. These values are of the same order of magnitude as
those of other self made MFM tips~\cite{KC97,LKCDW99}, and several
orders of magnitude larger than the effective magnetic moment of
commercial tips, where a $\approx50$~nm layer of CoCr is usually used.

\begin{figure}
\includegraphics[width=\columnwidth]{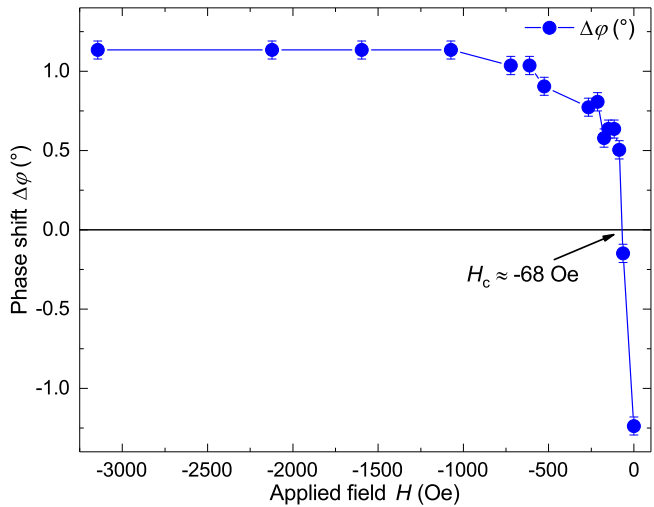}
\caption{\label{fig:fig7} Phase shift $\Updelta\varphi$ as a function of
  magnetic field $H$ applied in the opposite direction
  of tip magnetization, at \text{$T=4.2$~K.}}
\end{figure}

In order to obtain the coercive field of the magnetic Akiyama tip, a
magnetic field of $7$~T was applied prior to the measurements, in
order to magnetize the tip in $z$-direction. The field was turned off,
and as before, the $y$ axis was fixed at $y=0$ in order to scan the
center line across the loop. Several scans have been measured and the
phase signal was averaged. After each measurement the oscillation of
the tip was turned off, and a magnetic field was applied in the
opposite direction, and turned off again before measurement. The
results can be seen in \figurename~\ref{fig:fig7}.
A coercive field of $\approx68$~Oe was found and the magnetic
saturation sets in at field larger than $1000$~Oe.

\section{Conclusion}

FEBID of Co has been successfully used to produce magnetic Akiyama
tips suitable for MFM measurements at 300~K as well as at low
temperatures. Using FEBID, ferromagnetic Co can be deposited in
sufficient amount on Akiyama tips and other standard AFM probes, in
any desired shape and size. The results can be very well described
using standard theories. Akiyama tips are sensitive to electrostatic
interactions, however, this should not be a concern when measuring
ferromagnetic samples. Such self made magnetic tips can be used for
MFM measurements with all the advantages of such self-oscillating and
self-sensing tips, such as temperature dependent or light-sensitive
measurements.

\begin{acknowledgments}

This work was supported by the European cooperation in science and
technology (COST) through COST Action CM1301, ``Chemistry for
Electron-Induced Nanofabrication'' (CELINA), and by the DFG through
the Collaborative Research Center SFB 762 ``Functionality of Oxide
Interfaces''.

\end{acknowledgments}
\bibliography{bibliography}

\end{document}